\documentclass[twocolumn,epsfig,floats,showpacs]{revtex4}
\usepackage{graphicx}
\usepackage{color}

\usepackage{fancyhdr}
\fancypagestyle{titlepage}{
\fancyhead[R]{J. Phys. Chem. C \textbf{123}, 9688 (2019)}}
\begin{document}

\title{Vibrational Effects in X-ray Absorption Spectra of 2D Layered Materials}

\author{W. Olovsson$^{1}$, T. Mizoguchi$^{2}$, M. Magnuson$^{1}$, S. Kontur$^{3}$, O. Hellman$^{4,5}$, I. Tanaka$^{6}$, and C. Draxl$^{3,7}$}

\affiliation{$^1$Department of Physics, Chemistry and Biology (IFM), Link\"{o}ping University, Sweden}

\affiliation{$^2$Institute of Industrial Science, The University of Tokyo, 4-6-1 Komaba, Meguro-ku, Tokyo, Japan}

\affiliation{$^3$Physics Department and IRIS Adlershof, Humboldt-Universit\"{a}t zu Berlin, zum Gro{\ss}en Windkanal 6, 12489 Berlin, Germany}

\affiliation{$^4$Department of Physics, Boston College, Chestnut Hill, Massachusetts 02467, USA}

\affiliation{$^5$Department of Applied Physics and Materials Science, California Institute of Technology, Pasadena, California 91125, USA}

\affiliation{$^6$Department of Materials Science and Engineering, Kyoto University, Sakyo, Japan}

\affiliation{$^7$European Theoretical Spectroscopy Facility (ETSF)}

\date{\today}

\begin{abstract}
With the examples of the C $K$-edge in graphite and the B $K$-edge in hexagonal BN, we demonstrate the impact of vibrational coupling and lattice distortions on the X-ray absorption near-edge structure (XANES) in 2D layered materials. Theoretical XANES spectra are obtained by solving the Bethe-Salpeter equation of many-body perturbation theory, including excitonic effects through the correlated motion of core-hole and excited electron. 
We show that accounting for zero-point motion is important for the interpretation and understanding of the measured X-ray absorption fine structure in both materials, in particular for describing the $\sigma^*$-peak structure. 
\end{abstract}

\maketitle

\section{Introduction}
X-ray absorption near-edge structure (XANES) is a powerful technique for the characterization of materials. It is used to identify chemical environment and bonding of specific elements by monitoring the electronic transitions between core levels and unoccupied states. Likewise, electron energy-loss near-edge structure (ELNES) by transmission electron microscopy can provide almost identical information. To fully utilize this spectral information, a reliable theoretical analysis can provide the required insight into the nature of the observed excitations.

There is a long history of computing core-level spectra from first principles. The majority of calculations~\cite{Mizoguchi2010} is based on density-functional theory (DFT)~\cite{Kohn1999} utilizing the concept of a {\it core hole} in a supercell, also called the final-state approximation~\cite{Tanaka2005,Hebert2007}. Although this method has been successful in describing most of the spectral features, it turned out that sharp excitonic peaks appearing near the absorption edge, or intensity ratios, cannot be reproduced reliably. In such cases, one needs to go beyond the core-hole approximation and treat electron-hole interaction by solving the Bethe-Salpeter equation (BSE) of many-body perturbation theory~\cite{Olovsson2009a,Olovsson2009b,Olovsson2011}. In this work, we demonstrate with the examples of graphite and hexagonal boron nitride, that another important step beyond this methodology is required to understand the spectra of these layered systems.

Graphite and hexagonal boron nitride are archetypical 2D materials that have been intensively discussed in the literature. However, neither the carbon $K$-edge ($1s$) spectra in graphite nor that of boron in $h$-BN has been satisfactorily explained by {\it ab initio} theory. Due to their hexagonal layered structures their excitation spectra are characterized by in-plane and out-of-plane components. Both show a pronounced $\pi^*$-peak at the core edge, that stem from $p_z$ orbitals pointing in the direction perpendicular to the layers. The contribution of the in-plane $sp^2$ orbitals is recognized as the main origin of the $\sigma^*$-peak structure, located at roughly 6 eV above the $\pi^*$ core-edge. Detailed experimental investigations revealed that the $\sigma^*$-peak in both crystals exhibits a double-peak structure, labeled $\sigma_1^*$ and $\sigma_2^*$~\cite{Ma1993,Moscovici1996}. 

The C $K$-edge absorption spectra in graphite has been calculated by various theoretical methods, including a BSE scheme based on the pseudopotential approximation~\cite{Shirley1998}, a core-hole supercell method~\cite{Ahuja1996,Moreau2006} and the Mahan-Nozi\`{e}res-De Dominices (MND) method~\cite{Wessely2005}. None of them could resolve the double-peak structure. Similar double $\sigma^*$ peaks are observed in the boron $K$-edge of $h$-BN in XANES as well as ELNES experiments~\cite{Moscovici1996}. Like for graphite, previous DFT calculations did not obtain this striking feature~\cite{Tanaka1999}. On the other hand, pseudopotential-based BSE calculations~\cite{Carlisle1999} showed the presence of a "camel-back", however its origin was not clarified. In this work, we show that vibrational effects must be accounted for in order to understand the shape of the $\sigma^*$ region. 

The impact of phonons and temperature effects on electronic excitations is an emerging issue, however, there is no commonly accepted way of describing them from first principles. For example, the influence of symmetry-breaking effects from phonons or Jahn-Teller distortions, has been discussed in the literature for graphite and other materials~\cite{Ma1993,Skytt1994,Batson1993,Tinte2008,Gilmore2010,Harada2004,STanaka2005,Yasui2006}. Incorporating electron-phonon coupling into the Bethe-Salpeter equation,~\cite{Marini2008} the temperature-dependent optical spectra of silicon and $h$-BN were investigated. An alternative approach based on stochastic modeling based on the Williams-Lax theory was used to account for zero-point motions in the optical spectra of nano-diamonds.\cite{Giustino2013} More recent examples of core excitations concern the Mg $K$-edge in MgO~\cite{Nemausat2015}, and N $K$-edge in $h$-BN~\cite{Vinson2017}. Here, we approach the problem from two sides. First, we probe the sensitivity of the spectra to symmetry-breaking vibrational modes. Second, we apply an efficient statistical model~\cite{Shulumba2017,Shulumba2017b} to include the effect of electron-vibrational coupling on the near-edge structure.

\section{Methodology}
To obtain the X-ray absorption spectra, we solve the Bethe-Salpeter equation, using the open source code ~\cite{exciting,Sagmeister2009}, that has been successfully applied to $K$-edge excitations in other materials~\cite{Olovsson2013}. For a detailed description of the implementation, see Ref.~\cite{Sagmeister2009} and references therein. Being based on the all-electron full-potential linearized augmented plane-wave (FPLAPW) method, exciting gives access to the core region without further approximations than those inherent of the underlying exchange-correlation functional used for the DFT ground-state calculation. The latter is the generalized gradient approximation (GGA) in the PBE~\cite{Perdew1996} approach in our case. For both systems, experimental lattice constants are adopted. 

First, we consider a computationally efficient approach which can be used to explore the general effect of lattice distortions on the spectra. Namely, we limit our study to phonon modes at the $\Gamma$ point. For the $E_{\tt 2g}$ modes, the unit cells consist of four atoms in two planes. For graphite, we use an $11 \times 11 \times 3$ {\bf k}-mesh and include 13 states above the Fermi-level in the setup of the BSE Hamiltonian for the distorted systems. For $h$-BN a $9 \times 9 \times 3$ {\bf k}-mesh and 25 unoccupied states were sufficient to capture the absorption fine structure. 

Secondly, in order to probe the overall effect of lattice vibrations on the excitation spectra, we use an efficient stochastic sampling approach, see Refs.~\cite{Shulumba2017,Shulumba2017b} and references therein. We generate a set of structures to sample a canonical ensemble, averaging over the amplitude of each phonon mode:
\begin{equation}
	\langle A_{is} \rangle = \sqrt{\frac{\hbar (2n_s+1) }{2 m_i \omega_s}}.
\end{equation}

Given these amplitudes, supercells were constructed with the atomic positions given by
\begin{equation}
u_i = \sum_{s} Q_{is} \langle A_{is} \rangle \sqrt{-2\ln \xi_1}\sin 2\pi\xi_2,
\end{equation}
where, $0<\xi<1$ are uniform random numbers, $\omega$ the frequency, and $Q$ the eigenvector of mode $s$. The temperature enters via the Bose occupation factor $n_s$. Here, the supercells consist of 16 atoms placed in two planes. Fifty structures were used for the sampling of graphite and 100 structures for $h$-BN. For graphite, we use a $4 \times 4 \times 1$ {\bf k}-mesh and include 50 states above the Fermi-level in the setup of the BSE Hamiltonian. For $h$-BN the same {\bf k}-mesh and 80 unoccupied states were sufficient to capture the absorption fine structure.

For comparison with experiment, a Gaussian broadening of 0.2 eV full width at half maximum is applied to the spectra, which are aligned at the $\pi^*$ peak by a rigid energy shift of the DFT energies.

\section{Results and discussion}
\begin{figure}[t]
\includegraphics[width=70mm]{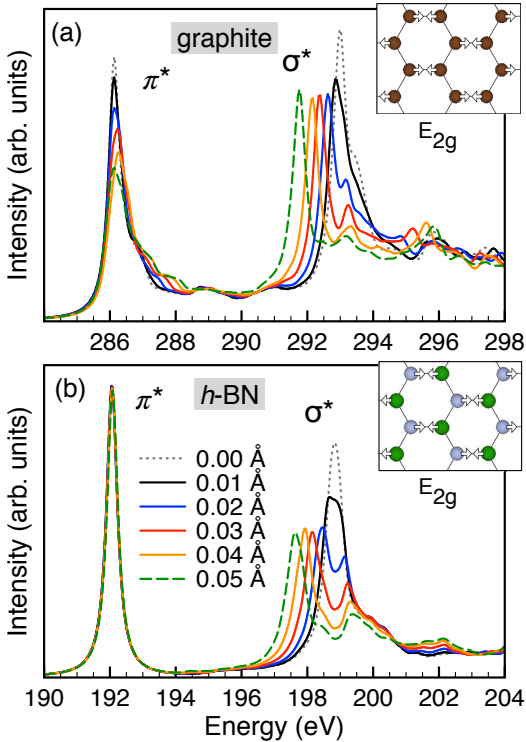}  
\vspace{0.2cm} 
\caption[] {Calculated core-level spectra from the solution of the BSE for a) graphite and b) $h$-BN, accounting for different displacements $\delta = 0.01$ to $0.05$ \AA\ according to the $E_{\tt 2g}$ phonon mode. The insets show the corresponding displacement patterns.}
\label{BSE}
\end{figure}

To consider the effect of distortions on the absorption spectra, we first limit our study to the $\Gamma$ point modes. We find that the $E_{\tt 2g}$ and $E_{\tt 1u}$ vibrations have the most significant effects on the spectra for both graphite and $h$-BN, and lead to very similar results. 
Other phonon modes show only smaller effects, in particular for out-of-plane movements of the atoms, as compared with the unperturbed lattice. The possible effect on the spectra can be recognized already from the respective unoccupied density of states for the cells.
Both $E_{\tt 2g}$ and $E_{\tt 1u}$ modes change the in-plane bond lengths as evident from the eigenvectors shown in the inset of Fig.~\ref{BSE}. The calculated BSE spectra for different vibrational amplitudes between 0.01 to 0.05 \AA, along the $E_{\tt 2g}$ phonon mode are shown in Fig.~\ref{BSE}. The first $\sigma^*$ peak is found to shift almost proportional to $\delta$. Most important, the $\sigma^*$ peak clearly splits into two, already at a displacement of 0.02 \AA. To demonstrate, that this is indeed important at temperatures where experimental spectra are typically recorded, we have computed the root mean square atomic displacement, $\delta_{\rm rms}$, as a function of temperature for the $E_{\tt 2g}$ and $E_{\tt 1u}$ phonon modes. We find that zero-point vibrations are dominating up to $\sim$500 K owing to the high phonon frequency modes of 47.4 THz for graphite and 40.2 THz for $h$-BN (not shown).
Without considering anharmonic effects, $\delta_{\rm rms}$ is around 0.03 \AA\ at RT and below. Only at extremely high temperatures, the quantum-mechanical displacement converges to the classical limit.
The great sensitivity of the spectra to atomic vibrations, already present by zero-point motion, also holds true for $h$-BN.

\begin{figure}[t]
\includegraphics[width=70mm]{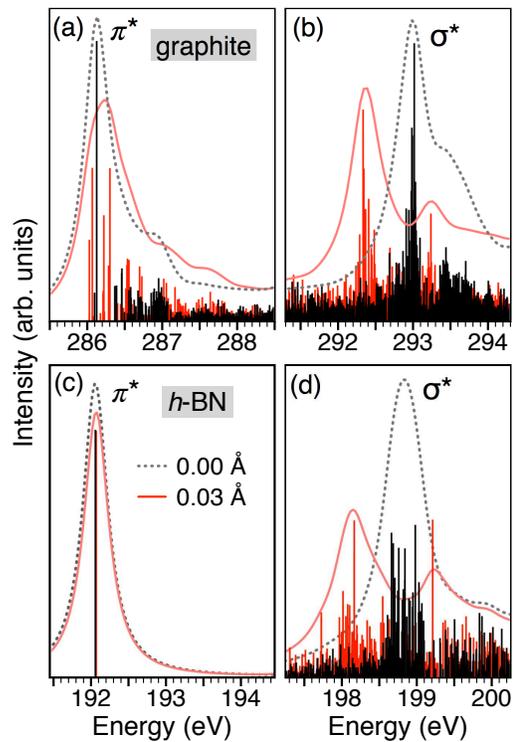}
\vspace{0.2cm} 
\caption[] {Relative oscillator strengths for the excitations contributing to the a) $\pi^*$ and b) $\sigma^*$ peak for the C $K$-edge in graphite and for the c) $\pi^*$ and d) $\sigma^*$ peak for the B $K$-edge in $h$-BN. Black lines correspond to the equilibrium positions and red lines to a displacement of $\delta = 0.03$ \AA\ in the $E_{\tt 2g}$ mode. The broadened spectra are indicated by dashed black lines.}
\label{OS}
\end{figure}

\begin{figure}[t]
\includegraphics[width=70mm]{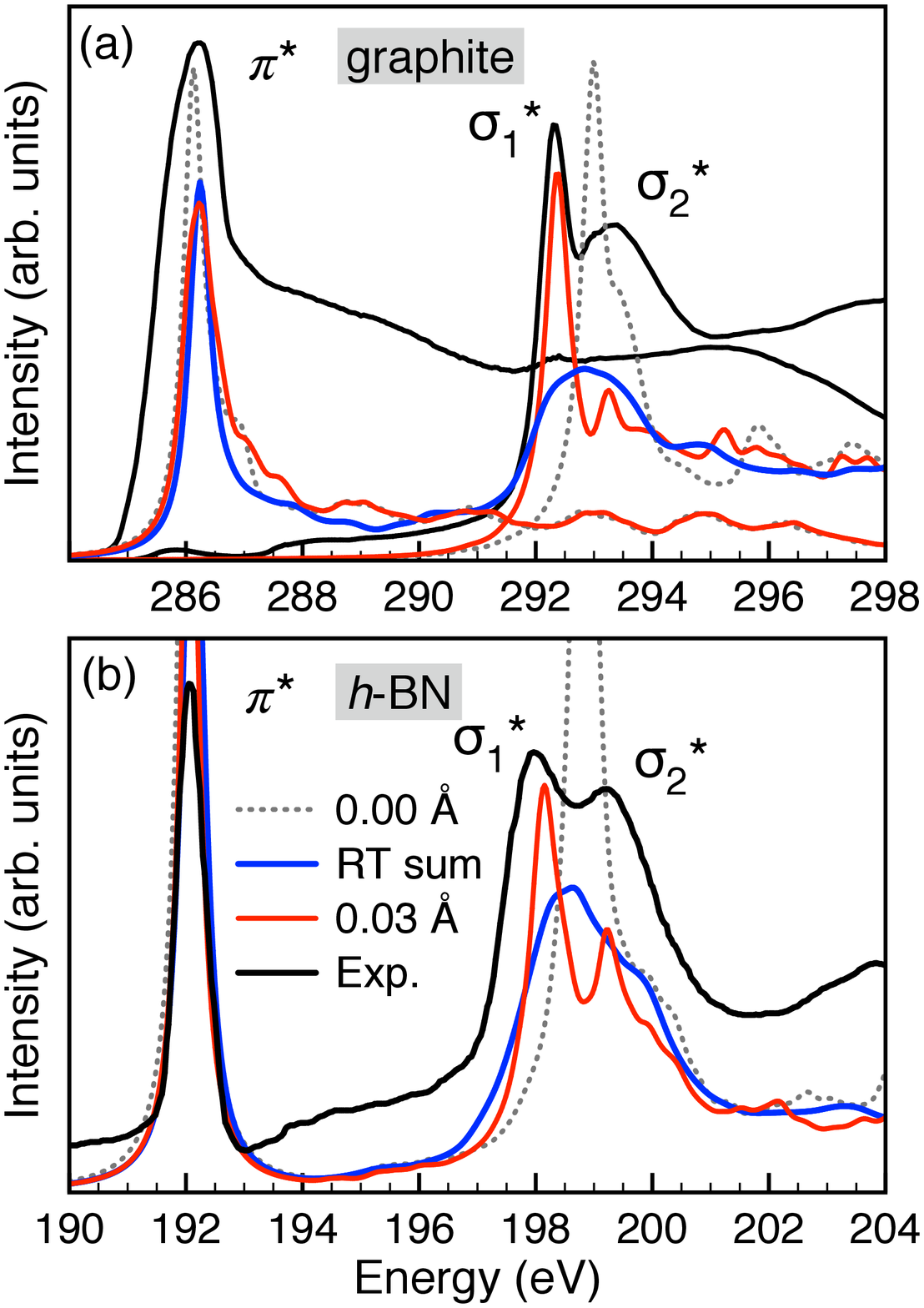}
\vspace{0.2cm} 
\caption[] {Experimental XANES (black full lines) and calculated BSE spectra for a) the carbon $K$-edge in graphite and b) the B $K$-edge in $h$-BN for the equilibrium geometry (gray dotted lines), the RT average (dark blue lines) and for displacements of $\delta = 0.03$ \AA\ (red lines) according to the $E_{\tt 2g}$ phonon mode.}
\label{XANES}
\end{figure}

In Fig.~\ref{OS} we show the single excitations contributing to the $\sigma^*$ and $\pi^*$ peak structures as obtained from the BSE calculations for the equilibrium geometry (black lines) and displacements according to the $E_{\tt 2g}$ phonon modes with $\delta$ = 0.03 \AA\ (red lines) for both materials. Many excitations with low oscillator strength between the $\pi^*$ and $\sigma^*$ peaks can be seen for graphite, but not for $h$-BN $\pi^*$. At the equilibrium structure, the core-edge in graphite consists of two strongly bound core excitons. More complicated excitonic features are found for the displaced geometries, characterized by an increasing number of excitations and redistribution of oscillator strength. In $h$-BN, the $\pi^*$ core-edge consists of a single strongly bound core exciton, which is practically not affected by the symmetry-breaking in-plane modes. Here also a slightly more strongly bound core exciton exists, but it has vanishing oscillator strength due to its $s$-orbital character. For both systems, the $\sigma^*$-region can be described as a mixture of different excitations, whose main features are several strongly bound core excitons with high oscillator strengths. In particular, the $\sigma^*_2$ structure in $h$-BN appears as mainly due to a strong single excitation, while several excitations are observed for graphite.

In Fig.~\ref{XANES}, we compare the calculated BSE results, i.e.\ the room temperature (RT) average (dark blue lines) -- further discussed below, the ones representing an $E_{\tt 2g}$ vibration with $\delta$ = 0.03 \AA\ (red lines), as well as the one for the equilibrium structure (gray dotted lines), with the experimental XANES spectra (black lines).
For graphite, the respective out-of-plane and in-plane contributions are shown separately. The experimental spectra were obtained for a highly oriented pyrolytic graphite (HOPG) sample of high purity manufactured by chemical vapor deposition (CVD) and cleaved to obtain a fresh surface. The measurement was performed at 300 K and $\sim$ 1$\times10^{-7}$ Pa at the undulator beamline I511-3 on the MAX II ring of the MAX IV Laboratory (Lund University, Sweden)~\cite{Magnuson2012b}. The energy resolution at the C $1s$ edge of the beamline monochromator was 0.1 eV. The spectra were recorded at 15$^o$ (along the $c$-axis, near perpendicular to the basal $ab$ plane) and 90$^o$ (normal, parallel to the basal $ab$-plane) incidence angles and normalized by the step edge below and far above the absorption thresholds. The experimental data for the B $K$-edge in $h$-BN are taken from Ref.~\cite{Li1996}.

We recall here, that the upper part of the X-ray absorption spectrum of graphite has so far been ambiguously interpreted, partly owing to the fact that first-principles studies~\cite{Ahuja1996,Shirley1998,Wessely2005,Moreau2006} could not reproduce the $\sigma^{*}_{2}$ peak. The feature was early on attributed to vibronic coupling by Ma {\it et al.}~\cite{Ma1993}, arguing for strong vibrational effects in diamond and graphite based on X-ray emission spectra. Symmetry breaking by vibrations in graphite was put forward by Harada {\it et al.}\ by model calculations of resonant X-ray emission~\cite{Harada2004,STanaka2005,Yasui2006}. In contrast, Br\"{u}hwiler {\it et al.}\ interpreted $\sigma_1^*$ as an excitonic feature, in line with Ref.~\cite{Ma1993}, and $\sigma^{*}_{2}$ as a delocalized band-like contribution~\cite{Bruhwiler1995}. Also, delocalized $sp^2$ orbitals without influence of the core-hole, i.e., an initial-state effect, was suggested~\cite{Wessely2005} as its origin. 

For both materials, we find that including the effect of the in-plane phonon modes at the $\Gamma$ point, as seen in Fig.~\ref{XANES}, essentially reproduce the double peak structures corresponding to the $\sigma^*$ peak observed in experiment. In the case of graphite, there is an effective widening of the $\sigma^*$ peak region into the measured fine structure with the $\sigma_1^*$ and $\sigma_2^*$ peaks~\cite{Ma1993,Moscovici1996}. A similar trend is observed for the B $K$-edge in $h$-BN, reproducing the characteristic camel-back like feature~\cite{Li1996}. A difference between these theoretical results of the two systems including the lattice distortion effect, is that the intensity and the shape of the $\pi^*$ peak display virtually no change in $h$-BN, as opposite to graphite.

\begin{figure}[!t]
\includegraphics[width=70mm]{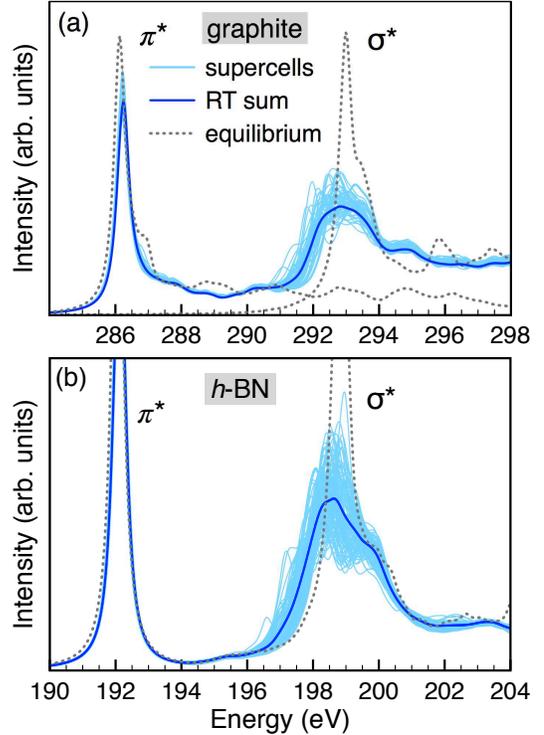}  
\vspace{0.2cm} 
\caption[] {Calculated core-level spectra from the solution of the BSE for a) graphite and b) $h$-BN, with results from the set of selected supercells (light blue lines) and the corresponding RT average (dark blue lines). For comparison, the spectra of the equilibrium structure is shown (gray dotted lines).
}
\label{ALLPH}
\end{figure}

We consider the spectra at room temperature (RT) by sampling over the canonical ensemble as described in the Methodology section. The result of the corresponding BSE calculations are shown in Fig.~\ref{ALLPH} with individual spectra for the supercells (light blue lines) compared with the average sum at RT (dark blue lines) and equilibrium (gray dotted lines). It is clear that the $\sigma^*$-region is significantly affected with a shift towards lower energy and a redistribution of intensity. In the case of graphite the sharp peak is much reduced into a broadened shape, while $h$-BN shows features closer to a double-peak structure. On the other hand, the positions of the $\pi^*$ peaks are almost the same, although with a slight effect at the C $K$-edge.
In general, the effect of the averaging process is a reduction of the dominant spectral features, originating from the in-plane phonon modes, as observed before. It remains unresolved though why these modes play a dominant role as one may conclude from the comparison with experiment. This observation points towards other mechanisms at play, e.g.\ core-hole life time effects, which are not included in the present modeling. Note added in proof: the recent work of Karsai {\it et al.}~\cite{Karsai2018} use the supercell and core-hole method in a related approach to obtain the double-peak structure in h-BN.

\section{Conclusions}
In summary, we have demonstrated the importance of including vibrational effects for XANES/ELNES spectra of the C $K$-edge in graphite and B $K$-edge in $h$-BN. We anticipate that zero-point motions and lattice symmetry breaking can be important for many other materials. Graphite and hexagonal boron nitride are archetypical layered structures, which share many features with the related 2D materials of graphene and BN monolayers. Thus we expect similar behavior also in these systems. Generally, we point to low-dimensional structures, where electron-phonon coupling is typically enhanced, and particularly to materials with light atoms, that exhibit high vibrational frequencies. Since we expect vibrational effects to be visible in a large temperature range, we encourage new temperature-dependent experiments on light-weight low-dimensional materials.

\section{Acknowledgement}
J.J.\ Rehr is acknowledged for providing access to the {\tt OCEAN} program~\cite{OCEAN}, used for comparative tests.
W.O.\ acknowledges support from the Swedish Government Strategic Research Area in Materials Science on Functional Materials at Link\"{o}ping University (Faculty Grant SFO-Mat-LiU no.\ 2009 00971) and Knut and Alice Wallenbergs Foundation project Strong Field Physics and New States of Matter CoTXS (2014-2019).
We would like to thank the staff at MAX-IV Laboratory for experimental support and Dr.\ Atsushi Togo for valuable discussions on theory.
M.M.\ acknowledges financial support from the Swedish Energy Research (no.\ 43606-1) and the Carl Trygger Foundation (CTS16:303, CTS14:310). 
The calculations were carried out at the National Supercomputer Centre (NSC) at Link\"oping University, supported by SNIC.
Support for I.T.\ by JSPS KAKENHI 26630295 and 25106005, T.M.\ by JSPS KAKENHI 26249092, and  
C.D.\ by the Deutsche Forschungsgemeinschaft through SFB 658 and SFB 951, are acknowledged.

\end{document}